\documentstyle[11pt,newpasp,twoside,epsf]{article}
\def\edcomment#1{\iffalse\marginpar{\raggedright\sl#1\/}\else\relax\fi}
\reversemarginpar

\begin{document}
\title{Weak lensing study of Abell 2029}
 \author{Brice M\'enard}
\affil{Max-Planck-Institut f\"ur Astrophysik, P.O.~Box 1317,
  D--85741 Garching, Germany}
\author{Thomas Erben}
\affil{Institut f\"ur Astrophysik und Extraterrestrische Forschung, 
Auf dem H\"ugel 71, D--53121 Bonn, Germany}
\author{Yannick Mellier}
\affil{Institut d'Astrophysique de Paris, 98 bis Bld Arago, F--75014
  Paris, France}

\begin{abstract}
Abell 2029 is one of the most studied clusters due to its proximity
(z=0.07), its strong X-ray brightness and its giant cD galaxy which is
one of the biggest stellar aggregates we know. We present here the
first weak lensing mass reconstruction of this cluster made from a
deep I-band image of $28.5'\times 28.5'$ centered on the cluster cD
galaxy. This preliminary result allows us already to show the shape
similarities between the cD galaxy and the cluster itself, suggesting
that they form actually a single structure.\\ We find a lower estimate
of the total mass of $1.8\times10^{14} h^{-1}\;M_\odot$ within a
radius of $0.3\;h^{-1}$Mpc.  We finally compute the mass-to-cD-light
ratio and its evolution as a function of scale.
\end{abstract}

\vspace{-0.75cm}
\section*{Introduction}

Abell 2029 is a well-known cluster, intensively studied in optical,
X-ray, infrared and radio. Its particularity is to host the biggest cD
galaxy ever observed.  These cD galaxies seem to be often found in
rich and relaxed clusters.  Investigating the relationship between the
diffuse light of these bright objects and the total mass of their
cluster might be a useful probe of the properties of galaxy clusters.
We present here the first weak lensing analysis of Abell 2029 which is
a preliminary result of a future detailed study.

\section{Weak lensing analysis}

The low redshift of this cluster is welcome for X-ray and optical
analyses; however the geometry of a very close lens and very distant
sources is in general a disadvantage for the lensing efficiency.
Therefore, a noisy mass reconstruction has to be expected in our study
compared to more distant clusters.  Furthermore, the cluster galaxies
in low redshift clusters have a large angular extend so that the
number density of background galaxies close to the cluster center is
small.  On the other hand, the mass computed for low redshift lenses
is not very sensitive to the exact source redshifts of the background
galaxies used for the reconstruction.

\begin{figure}
\begin{center}
\plottwo{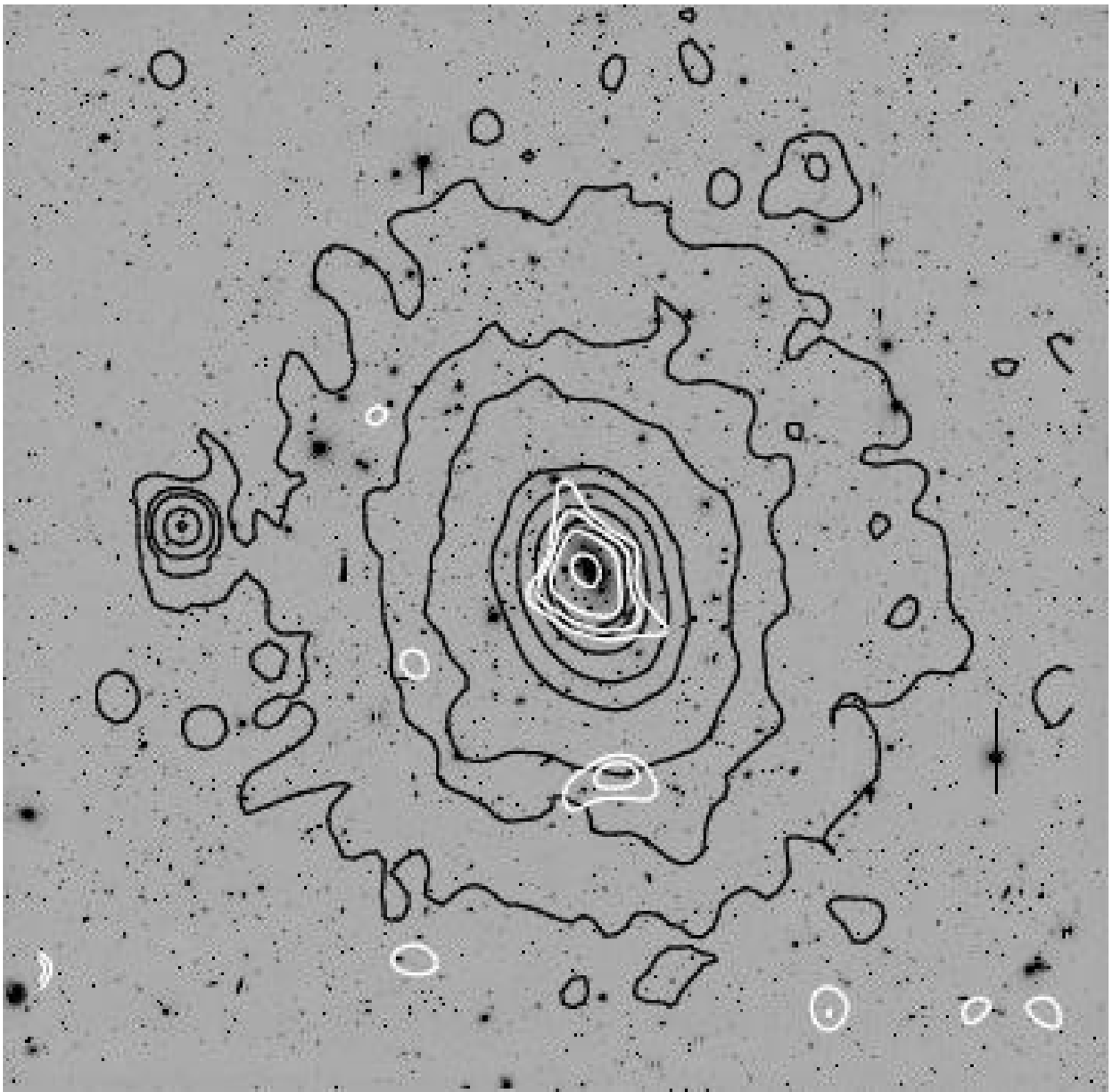}{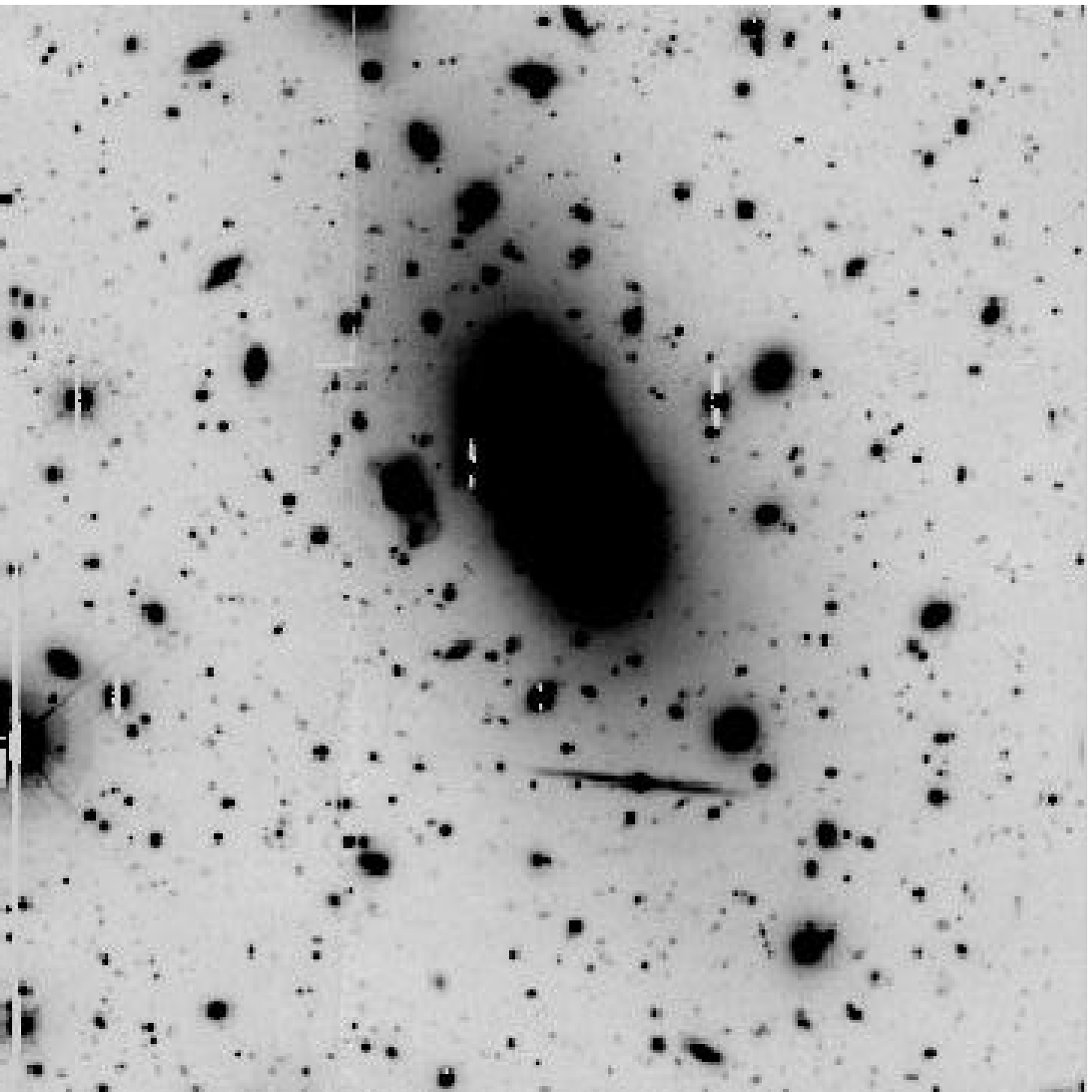}
\caption{Left panel : projected-mass map from weak lensing 
($\kappa=0.12$, 0.15, 0.2 and 0.3 ; in white) superposed on X-ray 
contours from ROSAT (in black). The scale of this cutout is 25 arcmin
on each side.
Right panel : close up near the center of the cD in the I-band
image. We show the innermost 1.5 arcmin of the mosaic.}
\vspace{-0.5cm}
\end{center}
\label{mass_map}
\end{figure}
Our data were obtained in the I-band under subarcsec seeing conditions
with the UH8K mosaic camera at the Canada-France-Hawaii telescope.
The UH8K consists of eight 2k$\times$4k chips arranged in a mosaic of
two rows each containing 4 CCDs. The total field-of-view is
28$'\times$28$'$ with a pixel scale of 0.206$''$. For our analysis we
did not use the upper right chip of the mosaic suffering from charge
transfer efficiency problems.  The total integration time for A2029
was 9600 sec. consisting of 8 individual exposures taken with a dither
pattern spanning about 15 arcsec to allow us to fill the gaps between
individual CCDs.  The exact details of the data processing will be
published elsewhere.

From the final coadded image we extract an object catalogue with
positions, sizes, ellipticities and magnitude.  Since we do not have
any color information we select the background galaxies by comparing
the magnitude and size distributions of the galaxies in the center of
the image (at the cluster-center location) and at the borders. The
overdensity in the former distribution indicates in a statistical
sense the magnitude and size of the cluster members.  We then define
our background galaxy catalogue by considering the corresponding
smaller and fainter galaxies. As a result of this procedure, a number
of background galaxies will not be included in our sample and a number
of foreground galaxies might be included as background ones. This will
not affect the shape of the mass contours but will lower the amplitude
of the measured gravitational shear and allow us to estimate only a
lower bound of the cluster mass.

The final catalogue contains $9\;314$ background galaxies
($\sim$15/arcmin$^2$).  To obtain final estimates of the galaxy
shapes, the raw galaxy shapes were corrected for PSF effects with the
method described by Kaiser, et al. (1995) and moficiations described
in Hoekstra et al. (1998) and Erben et al (2001).  This correction was
applied separately on each CCD and the density of approximately one
hundred stars per chip allowed an accurate PSF-correction mapping.

\subsection{Mass distribution}

The catalogue allows us to estimate a smoothed shear map from that we
can perform a weak lensing mass reconstruction. The left panel of
Fig. 1 shows the resulting $\kappa$-map, i.e. the projected mass map,
from the shear smoothed with a Gaussian filter with
$\sigma=68\arcsec$.  The map clearly shows the presence of the cluster
as well as several additional peaks of lower amplitude.  In order to
quantify the significance of each peak in our field we used the
$M_{\rm ap}$ statistics introduced by Schneider (1996).  With a filter
scale of 4.12 arcmin, the cluster shows up as a 6.5-$\sigma$
detection. Besides the main cluster, we find the secondary peaks to be
likely noise features given their detection levels (Van Waerbeke
1999).  The mass reconstruction clearly shows that the center of the
mass coincides with the center of the cD galaxy and that the mass is
elongated in the same direction as the light.  We clearly see the
strong correlation between the cD light and the mass distribution at
the cluster center.

Next, we have computed the mean tangential shear around the cluster
center in independent circular bins (see Fig. 2).  Hereby we chose as
center the position of the peak in the $M_{\rm ap}$ statistics with
the smoothing scale of 4.12 arcmin where the cluster is detected at
the highest significance. The massive cluster clearly causes a
significant shear signal up to the borders of our data field.  The
error bars are estimated from the dispersion of the cross shear
component that is not influenced by gravitational lensing.  The
profile is well fitted by an isothermal sphere model giving a best fit
velocity dispersion of 842 km s$^{-1}$.  As we have seen before, this
value has to be interpreted as a lower limit of the real one.

The X-ray contours that are also shown in Fig. 1 are centered on the
cD galaxy and show the same orientation as the mass and the light
distribution.  The significant X-ray emission of the cluster goes even
beyond the field of our optical image.  The similar distributions of
the cD light, the X-ray gas and the total mass strongly suggest that
the cD is not an isolated object, but an aggregate of stars orbiting
in the cluster potential.
 
\begin{figure}
\label{shear_kappa}
  \plottwo{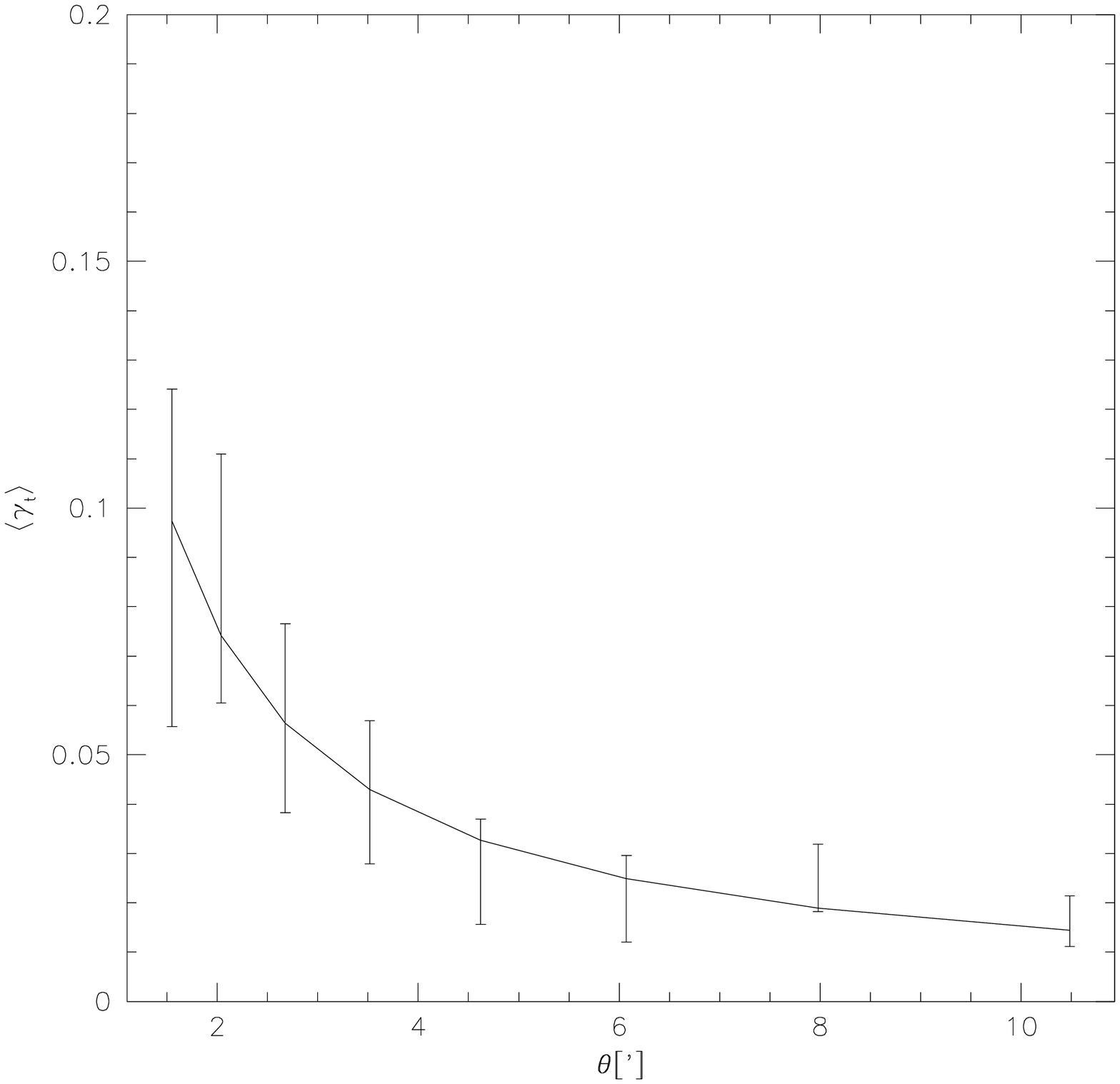}{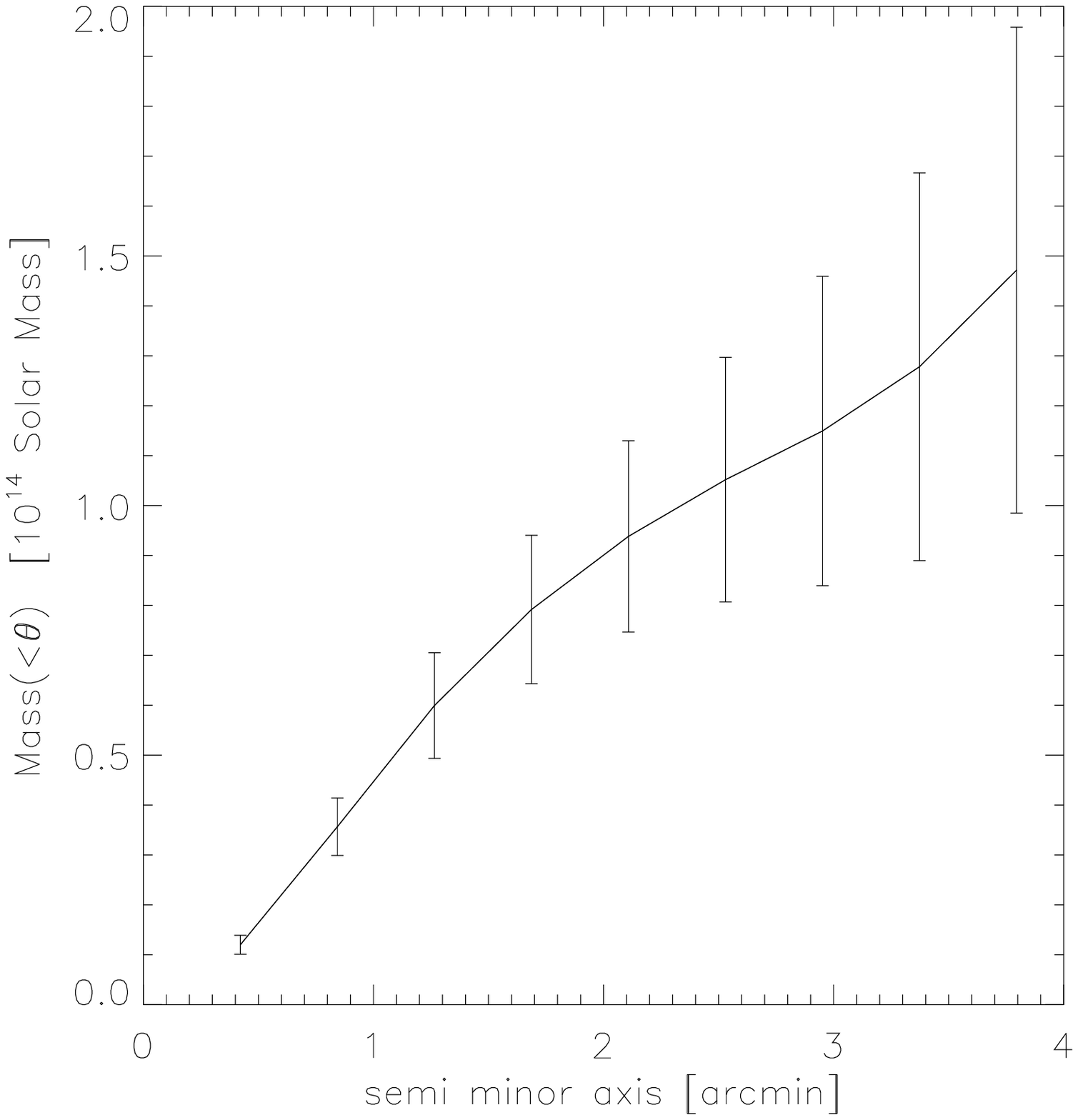} 
\caption{Left panel : shear profile measured in independent circular bins and 
fitted by an isothermal sphere profile. 
Right panel : projected-mass measured in ellipses of
1:2 axis ratio and aligned with the cD light.}
\end{figure}

The quantitative estimation of the mass requires breaking the
so-called mass-sheet degeneracy by knowing or assuming the value of
the mass at a given angular position. In order to do so, we assume no
mass at the borders of the field, i.e. far from the cluster center and
force the average $\kappa$-value at these locations to be
zero. However, since the X-ray map and the shear profile still
indicate the presence of some mass at a distance of $17\arcmin$ from
the cluster center, this further constrains our mass estimate to be a
lower limit only.

Using an elliptical bin with the same orientation and shape as the cD
light, we find a mass of $1.8\times10^{14} h^{-1}\;M_\odot$ within a
semi-minor axis of $0.3\;h^{-1}$Mpc.
\section{Light distribution and M/L ratio}
Uson et al. (1991) studied in detail the diffuse light of A2029 in
R-band.  They found that in order to trace this light component out to
large radii, correcting the data for the contamination of extended
halos of bright stars is necessary.  They applied these corrections by
accurately measuring their Point Spread Function (PSF) around very
bright stars in other fields and then substracting this PSF around the
bright stars present in the cluster field.  This allowed the detection
of the extended halo of the cD galaxy out to a distance of
425$\;h^{-1}$ kpc, measured as $d=\sqrt{r_{\rm min}r_{\rm max}}$.  The
cD envelope appears with a constant eccentricity and follows a de
Vaucouleurs profile over the whole range of detection (see Fig. 3).
The integrated luminosity of the cD galaxy with this halo is 5
$\times\,10^{11}h^{-2}\,L_\odot$ (R-band).

In order to study the relationship between the cD light and the
cluster mass, we have computed the $M/L_{\rm cD}$ ratio in R-band as a
function of scale (see Fig. 3).  This lower limit evolves from 100 to
300 M$_\odot/$L$_\odot$ in the innermost 4 arcminutes.

\begin{figure}
  \plottwo{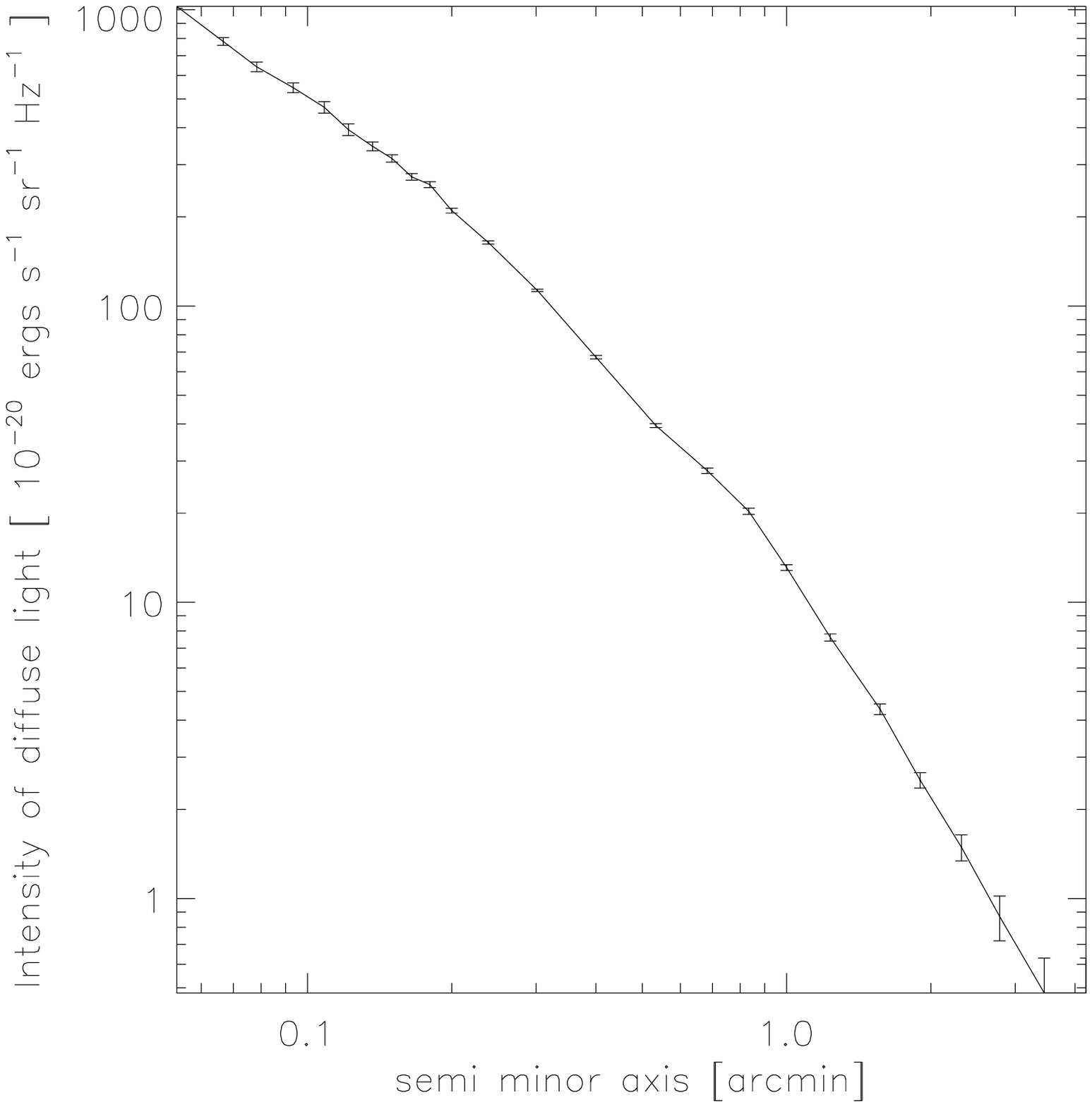}{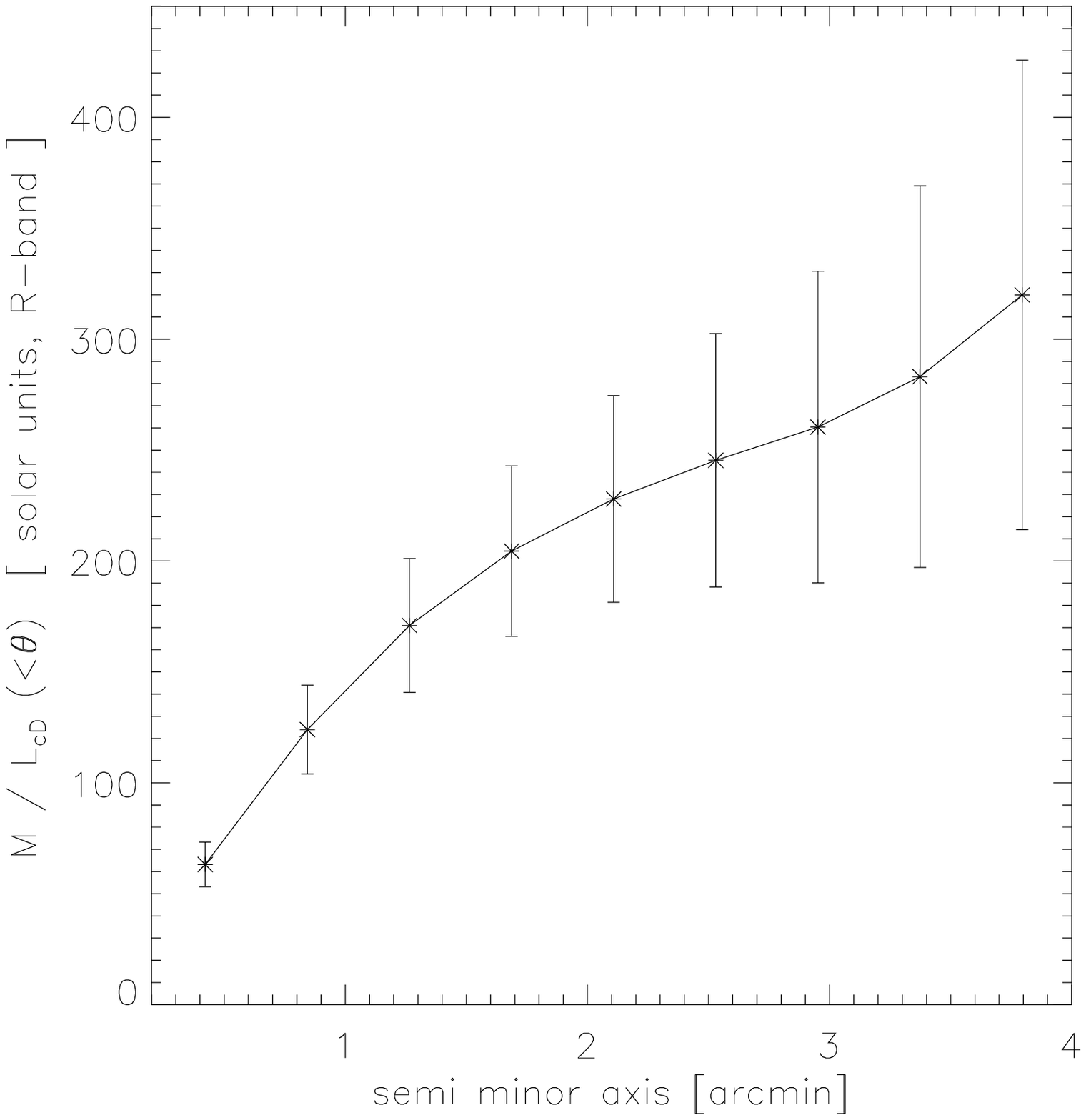}
\caption{R-band light profile of the cD galaxy, from Uson et al. (1991) and 
corresponding lower estimate of the M/L$_{cD}$ ratio as a function of semi-major axis.}
\label{M_L}
\end{figure}
\section{Conclusion and Outlook}
\label{Conclusion} 

We have presented the first weak lensing analysis of Abell 2029.  The
very low redshift of the cluster imposes a low lensing efficiency and
therefore a noisy mass reconstruction compared to more distant
clusters.  The available data allowed only lower mass estimates of the
cluster since we can not properly select the background galaxies.
However we can compute the shape of the mass contours of the cluster.
They indicate that the cD galaxy lies in the center of the
gravitational potential of the cluster and we find similar shape and
orientation for the distribution of the mass and cD light.  We have
seen that despite the large size of our field ($28.5' \times 28.5'$)
the mass can be traced up out to the border of the CCD which prevents
us to break the mass-sheet degeneracy.

Using the light distribution of the cD galaxy studied in detail by
Uson et al. (1991), we have computed the profile of the
mass-to-cD-light ratio, which reaches a value of 300
M$_\odot/$L$_\odot$ in R-band at 4 arcminutes.

These preliminary results of a more detailed future study can already
be used for comparing estimates made with other techniques.  Next we
plan to study this interesting cluster by using deeper images in two
colours with Megacam@CFHT.  Color information will allow us to obtain
a fair separation of the cluster galaxies and therefore a non biased
estimate of the cluster mass.  It will then be possible to compare the
light and mass distributions in great detail.  With a deep, 6 hour
I-band observation and the expected high number of background galaxies
we will be able to investigate the cluster mass out to larger scales
and thus allow a detailed comparison with existing X-ray maps and a
properly break the mass sheet degeneracy.

This work was supported by the TMR Network ``Gravitational Lensing:
New Constraints on Cosmology and the Distribution of Dark Matter'' of
the EC under contract No. ERBFMRX-CT97-0172, and by the Deutsche
Forschungsgemeinschaft under the project SCHN 342/3--1.


\end{document}